# WST - Widefield Spectroscopic Telescope: design of a new 12m class telescope dedicated to widefield Multi-object and Integral Field Spectroscopy

P. Dierickx[a], T. Travouillon[b], G. Gausachs[b], R. Bacon[a], C. Cudennec[a], I. Bryson[c], D. Lee[c],
J. Kragt[d], E. Muslimov[e&f], K. Dohlen[f], J. Kosmalski[g], J. Vernet[g], T. Lépine[h], P. Doel[i], D. Brooks[i]

[a]Centre de Recherche Astrophysique de Lyon, France ; [b]Australian National University, Australia ; [c]Science and Technology Facilities Council (STFC), United Kingdom ; [d]NOVA optical infrared instrumentation group at ASTRON, Netherlands ; [e]Dept. of Astrophysics, University of Oxford, United Kingdom ; [f]Laboratoire d'Astrophysique de Marseille, France ; [g]European Southern Observatory, International ; [h]Université Jean Monnet de Saint-Etienne, CNRS, Institut d'Optique Graduate School, Laboratoire Hubert Curien UMR 5516, France ;
[i]University College London, United Kingdom.

## ABSTRACT

The Wide-Field Spectroscopic Telescope (WST) is a concept for a 12-m class seeing-limited telescope providing two concentric fields of view for simultaneous Multi-Object Spectroscopy and Integral Field Spectroscopy. The specified wavelength range is 0.35-1.6 µm. The baseline optical design relies on a corrected Cassegrain solution feeding Multi-Object spectrographs through fibres, while the central area of the field is propagated down to a gravity-stable Integral Field Station housing 144 spectrographs. The Cassegrain corrector also provides for atmospheric dispersion compensation. All optical components are within commercially available dimensions. With a view to minimizing risks and costs, to the maximum possible extent the telescope relies on proven subsystem solutions. An exception is the tip-tilt secondary mirror, which would likely have to provide some rejection of wind shake. An iteration of the optical design is ongoing, with a view to mitigating the weaknesses of the first baseline design. The telescope would be wavefront-controlled on-sky at the common-path MOS focus. Controls in the IFS path will need to compensate for the effect of subsequent differentials – wavefront and line of sight. There is no shortage of degrees of freedom and metrology solution to do so. The size of the dome is driven by the Nasmyth footprint and the height of the pier, which houses the IFS station. The baseline assumption is that a VLT-like enclosure would provide suitable shielding and ventilation.

## 1. INTRODUCTION

WST is the product of a collaboration between European, British, Swiss and Australian institutes. It builds on early exploratory work for a massively multiplexed, 10-m class telescope[1]. A proposal to advance the phase design A has been submitted to the European Commission in March 2024 within the framework of the Horizon 2024 call. The proposed work would start with the baseline design developed over the last two years and described in this paper. A decision is expected by third quarter, 2024. The phase A would span years 2025 to 2027 and produce a costed proposal for detailed design and construction, for the European Southern Observatory to consider as its next major project after its Extremely Large Telescope[2] – the latter being under construction, with a planned first light by 2028. The eventual site for WST is assumed to be in Paranal area, with similar seismic, wind and turbulence conditions.

The phase A objective is to develop a point design of a peerless facility, *simultaneously* providing 3D spectroscopic capability *and* large-field multi-object spectroscopy. The design shall be supported by cost and schedule estimates, include an operational scheme, and respond to an integrated set of science objectives[3]. Top level requirements (Table 1) are still subject to consolidation but have already driven selection of a first baseline opto-mechanical solution. This design serves as a starting point; it was arrived at via a comprehensive trade-off. Focus will be on iterating some parameters to mitigate or eliminate known weaknesses.

A driving requirement is the simultaneous availability of two foci, with different optical properties. The ~f/3 field of view allocated to the Multi-Object Spectrographs (MOS) shall cover an area of not less than 2.5 squared degrees. At the same time, the telescope shall deliver a patrol field, up to 13 arc minutes diameter, within which any 3 x 3 arc minutes may be selected and sent to a gravity-invariant Integral Field Station (IFS) consisting of 144 identical spectrographs. Sampling at the MOS and IFS foci is specified in Table 1. The MOS and IFS instruments are described elsewhere; this paper focuses on the design of the telescope. Optical quality at the MOS focus shall be commensurate with fibre size, expected to be in the range 1.0 to 1.3 arc seconds, and good median seeing (~0.65 arc seconds Full Width at Half Maximum).

*Table 1 - Top level requirements, tentative*

| colspan="3" | Science / operation- driven requirements |
|---|---|---|
| *Parameter* | *Desired range / characteristic* | *Rationale* |
| Entrance pupil diameter | ~ 12-m | Collecting area twice larger than that of the Very Large Telescope |
| Vignetting | ≤ 12% | Idem |
| Field of view (FoV, area) | MOS   2.5 ≤ FoV ≤ 5 degrees squared<br>IFS    any 3 x 3 arc min² within a patrol field of 13 arc min diameter | Multiplexing<br>See [3] |
| MOS and IFS field of views simultaneously available | | See [3]; operational efficiency |
| Dead zone (obscured) between MOS and IFS fields shall be minimized, and in any case smaller than half the IFS patrol field diameter. | | See [3]; allow as-contiguous as possible IFS and MOS fields |
| Wavelength range | Required  0.37 – 1.0 μm (MOS & IFS)<br>Goal         0.35 – 1.6 μm (MOS only) | See [3] |
| Sampling | MOS   TBD, expected within 1.0-1.3 arc seconds<br>IFS    0.25 arc seconds | See [3] |
| Sky coverage | Full sky coverage up to z=60º<br>Goal up to z=70º | Operational efficiency, Targets of Opportunity |
| Optical quality (dia. 80% encircled energy) | MOS    ≤ 0.8 arc seconds<br>IFS     ≤ 0.2 arc seconds | Provisional guess; telescope only, as-designed |
| colspan="3" | Technological / design requirements |
| *Parameter* | *Desired range / characteristic* | *Rationale* |
| Gravity-invariant IFS station | | Size of the instrument, dimensional stability |
| IFS field de-rotation | Optical | Idem |
| Focal ratio – MOS | f/3 to f/3.5 | Fiber optical interface |
| Focal ratio – IFS | f/15 to f/50 | IFS optical interface |
| Field curvature | Minimise, preferably concentric to exit pupil, preferably convex as seen by an observer located post-focus. | Optical interfacing; MOS: fibers preferably orthogonal to focal surface and pointing towards exit pupil |
| Monolithic mirrors | Physical diameter ≤ 4.2-m | Availability on the market. Technological & procurement risk deemed too high beyond this limit. |
| Fast steering mirror(s) | Physical diameter ≤ 2.5-m | Risk; expected closed-loop bandwidth ~3-5 Hz |
| Reserved space behind primary mirror | ≥ 2.5-m | Maintenance (access), no light beams in heat dissipation volumes (primary mirror controls) |
| All inter-segments edges shall be visible from at least one of the foci | | On-sky phasing calibrations |

The strategy is to de-risk, to the maximum possible extent, the design of the telescope. At sub-system level, designs are constrained to commercially and operationally proven ones.

Assuming extensive re-use of ELT designs (e.g. segments, controls) and some relaxation of tolerances, for the telescope and enclosure a 10-years schedule for detailed design and construction would seem plausible. For reference, ESO's Very Large Telescope took 11 years from approval (April 1987) to first light (March 1998). We also note that by the time of WST detailed design and construction, there will be operational experience with ESO's significantly more demanding segmented, wavefront-controlled ELT, and most of its underlying industrial effort will have been completed[5].

## 2. SYSTEM DESIGN

We anticipate an alt-az design with segmented primary mirror and some significant level of wavefront control, merging VLT-like active optics and ELT- or Keck-like phasing control. Herein the term *wavefront control* refers to the set of functions, which are needed for the telescope to deliver both foci at the specified prescription. It includes dynamically controlling plate scale, line of sight (wavefront tip-tilt), and wavefront quality, simultaneously at both foci.

As the telescope is required to deliver seeing-limited images only, we do not anticipate fast, adaptive-like wavefront control except for the effect of wind buffeting on the structure and possibly on the primary mirror i.e., low order modes up to a few Hz at most. We assume that a VLT-like enclosure with adjustable louvers would allow to adjust to prevailing wind and local, induced turbulence conditions.

Early optical designs converged towards two families of solutions:

- Quad-mirror options with MOS fiber positioners either at Nasmyth or Cassegrain location, and an IFS field extracted from an intermediate focus, then propagated down to an IFS station located in the pier of the telescope. Atmospheric dispersion compensation is provided by way of 2 decentering lenses near the MOS focal surface.

- Corrected Cassegrain, with MOS fiber positioners and IFS field extractor located above the primary mirror, a Nasmyth-located 13 arc minutes intermediate focus, with an IFS sub-field selector and a coudé relay down to the IFS station located in the pier of the telescope. Atmospheric dispersion compensation is provided by the corrector lenses.

While delivering superb optical quality, for geometrical reasons the Quad options are field-limited to about 2.5 squared degrees by vignetting and imply several large and expensive 3 to 5-m class mirrors embedded in the altitude structure and at the location of the altitude bearing. Extraction for maintenance becomes challenging to say the least, and even drives the enclosure size. Besides, both MOS and IFS have large optics in their own, non-common paths. As a result, a coherent wavefront control scheme holding the two foci to specification in real time, with manageable cross-talk, becomes a risky and operationally complex proposition. Finally, the optical interface with MOS fibers is inconvenient as curvature of the MOS focal surface is opposite to the direction of the exit pupil – the focal surface as seen by an observer post-focus is concave.

A compact Cassegrain layout, with a three silica lenses corrector, has been selected as the most realistic option (Figure 1). It provides two concentric, simultaneously available fields – a 3.1 squared degree f/3 one (MOS) at a Cassegrain focus located above the primary mirror, and a 13 arcmin diameter f/35 one (IFS) in a Nasmyth configuration. Within the latter, any 3 x 3 arcmin$^2$ sub-field can be selected and propagated through a coudé relay, down to the gravity-stable IFS station located in the telescope pier. The IFS patrol field is picked by a small flat mirror (M3, 15cm) located just before the MOS focal surface. The dead zone separating the IFS and MOS fields i.e., the annular area lost to vignetting by the edges of this pick-up M3 mirror, is estimated at 3 arc minutes radial extension, but, pending mechanical implementation, this figure may be optimistic. Minimising the dead zone implies that this M3 pick-up mirror be located as close as possible to the MOS focal surface. This leads to an inevitable inconvenience: if the mirror is supported by spiders connecting to the altitude structure, the shadow of these spiders will rotate over the mechanically de-rotated MOS field. If it is supported through the MOS front-end assembly, it will either need to be mounted onto a ~1.3-m "fixed" window or be "de-de-rotated" in order to hold the IFS line of sight. We anticipate that the latter solution will prevail. Tolerances and bandwidth applicable to this "de-derotation" out to be modest, and the IFS path will need to hold its own exit pupil position and line of sight anyway.

The optical design has been optimised to cover the 0.35 – 1.6 μm wavelength range and produces an edge-of field image quality of 0.88 and 0.20 arcsec (80% encircled energy, diameter, worst case) at the MOS and IFS foci, respectively. As stated later herein, there is evidence that these figures can be improved. Atmospheric dispersion compensation is provided by combined decenters within the corrector, residual coma being taken care of by decentring the secondary mirror. All optics are within commercially available dimensions, including the 1.6-m largest lens of the refracting corrector. Designs are systematically sanity-checked for ghosts and pupil concentration.

MOS field curvature is concave towards the exit pupil, but not concentric to it, with the direction to the centre of curvature and the direction to the exit pupil deviating by typically 5-10 % of the image space full aperture angle (~D/f). The volumes allocated to the fibre-fed MOS instruments are located on the azimuth floor, for minimal fibre length.

Even though the telescope structure is compact (Figure 3), we anticipate that wind perturbation will be a significant contributor to the error budget. In the current baseline, with a diameter of 3.1-m the secondary mirror is deemed too large to allow for rapid (~3-5 Hz) tip-tilt control hence rejection of wind perturbation. At the time of writing of this article, solutions with a 2.4-m secondary mirror, lower aspheric departures in the refracting corrector, and better optical quality (MOS field, 80% encircled energy within ~0.70-0.80 arc seconds), have been identified, with no significant change of the overall layout. As far as the risk register is concerned, the reduction of the tip-tilt secondary mirror diameter is seen as a major improvement. Two options could be explored in phase A: either a ~50-mm, ~600 Kgs (glass only) moderately active meniscus or an ultra-lightweight, glass-ceramic, rigid mirror of similar mass. The target dimension becomes comparable to that of the 2.1 x 2.7-$m^2$ flat, field-stabilisation M5 mirror of the ELT[7]. The latter is made of sintered SiC, with a mass of about 400 Kgs. We expect WST performance specification (stroke, bandwidth, accuracy) to be less stringent and thus allow for less risky and more conventional mirror substrate options.

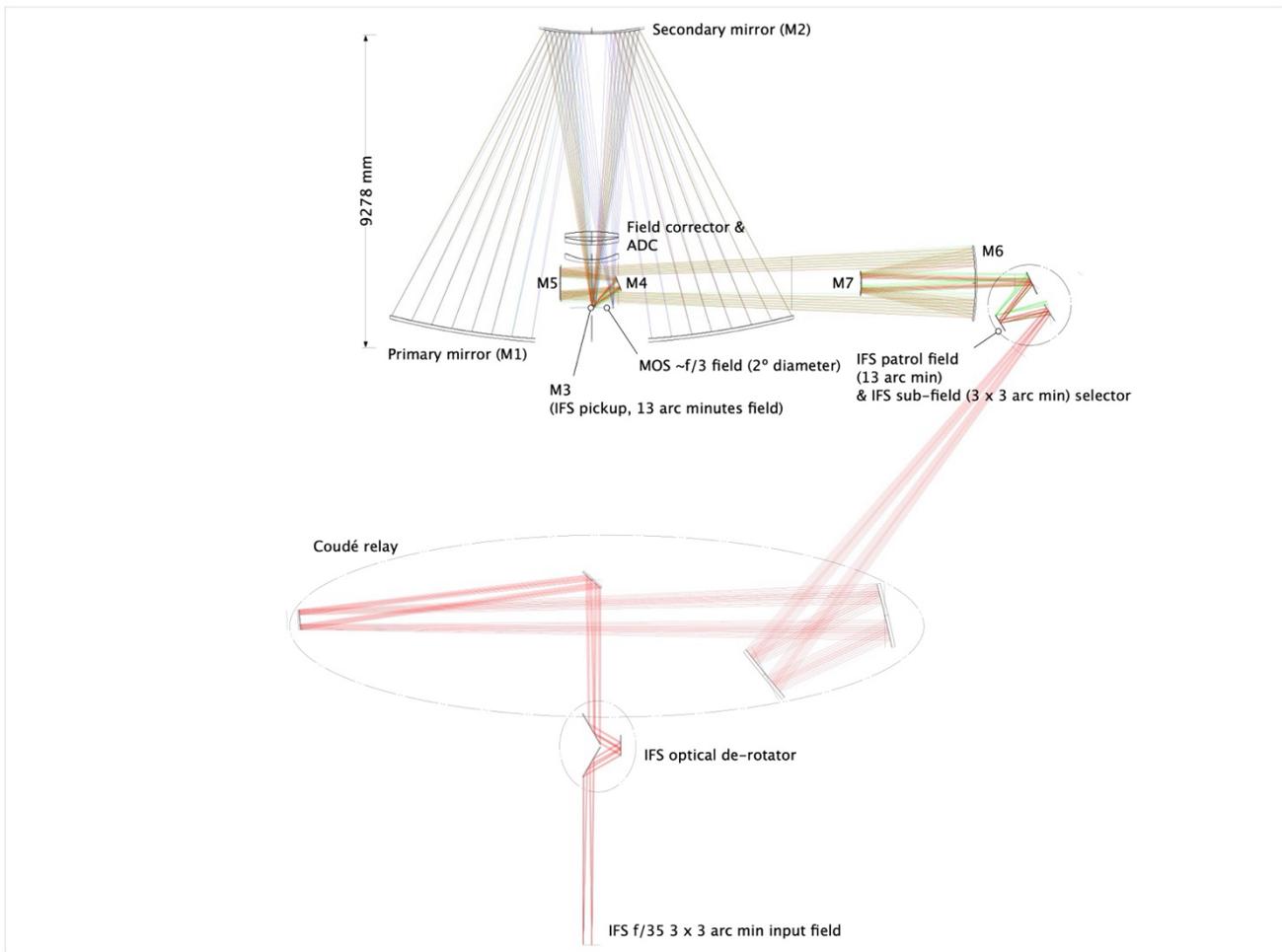

Figure 1 – WST optical layout, Cassegrain design, first baseline.

An evident weakness of the baseline is the number of optical surfaces down to the IFS station. Phase A will explore enhanced 0.37-1.0 μm coating options; to accommodate commercially available options, one of the objectives of the design iteration is to reduce the size of the largest coudé relay mirrors to 1.6-1.8-m.

Vignetting in the MOS focal surface is about 8% on-axis, increasing to 13% at 0.9º off-axis, and ~18% at the very edge of the field. The situation is less favourable for the IFS path, whereby central obstruction is driven by mirror M4, which sits in the beams sent sideways towards the Nasmyth station (Figure 1). The effect is nearly constant in the IFS field, and amounts to about 20% obscured area. The optical design effort is currently focusing on minimising this obscuration but geometrical constraints are such that we do not anticipate major gains with the current M3-M4-M5 arrangement. A promising alternative using off-axis conical mirrors is currently being explored. The overall layout remains broadly similar to that of the current baseline, thereby preserving the structural design to a large extent, and the central obscuration in the IFS path becomes identical to that of the MOS one.

The primary mirror has a total collecting area of 99 m$^2$ and consists of 78 segments. There are 13 distinct families of segments; the total number to be procured is therefore 91 units, including spares – a mere 10% of the ELT total production. Re-coatings and phasing runs would occur on a weekly to bi-weekly basis. The design of the telescope altitude structure includes a folding bridge crane for segment extraction and replacement. Setting aside their optical surface specification, the segments could be identical to those of the ELT[8], with the same supports and controls – even though stroke requirements for the position actuators will certainly be less challenging.

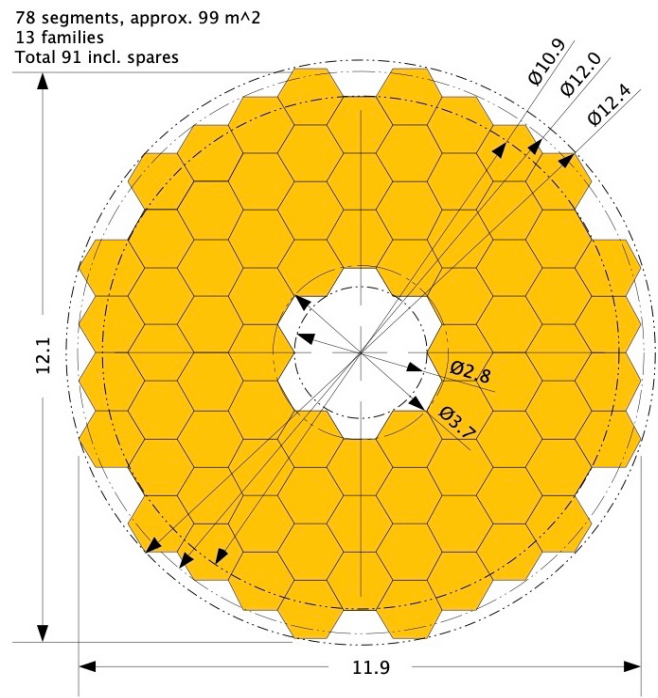

Figure 2 - WST Primary mirror geometry.

The guiding principle that will underlie the design of the wavefront control strategy is that the telescope shall be able to take care of itself i.e. it shall not need feedback from instruments to acquire target and wavefront. This does not mean that in operation, feedback from the instrument shall be *excluded*; it only means that on its own the telescope shall be able to deliver target and wavefront to the focal surfaces, within specification.

A critical advantage of the Cassegrain over the Quad design is that for all practical purpose the MOS optical path is also common path – setting aside the small, flat pickup mirror M3, located very close to the MOS focal surface. Cross-talk between wavefront control functions driven from the MOS focus and functions driven from the IFS one will therefore be limited. At most, IFS-driven wavefront control may send offload requests to the MOS-driven one. In a nutshell, target and wavefront can be acquired in the MOS focal surface, with IFS-specific differentials being managed in the IFS path. There is no shortage of potential degrees of freedom in the IFS path to do so, and the intermediate IFS field (13 arc minutes patrol field) is large enough to accommodate wavefront sensing. IFS spatial sampling will need to be commensurate with low frequencies only and we do not expect much limitation in terms of sky coverage. Neither do we expect large stroke requirements as most IFS-specific optical units are either small (15cm to 1.3-m for mirrors M3-4-5) or within relatively stable environments (Nasmyth station and coudé optics). Between the Nasmyth-located intermediate focus and the IFS station, we anticipate low perturbations – mostly thermal and runout of the optical de-rotator. This will imply either an internal metrology arrangement or on-sky sensing of tip-tilt and possibly focus. At f/35, plate scale and depth of focus are generous and we do not anticipate exceptionally demanding tolerances.

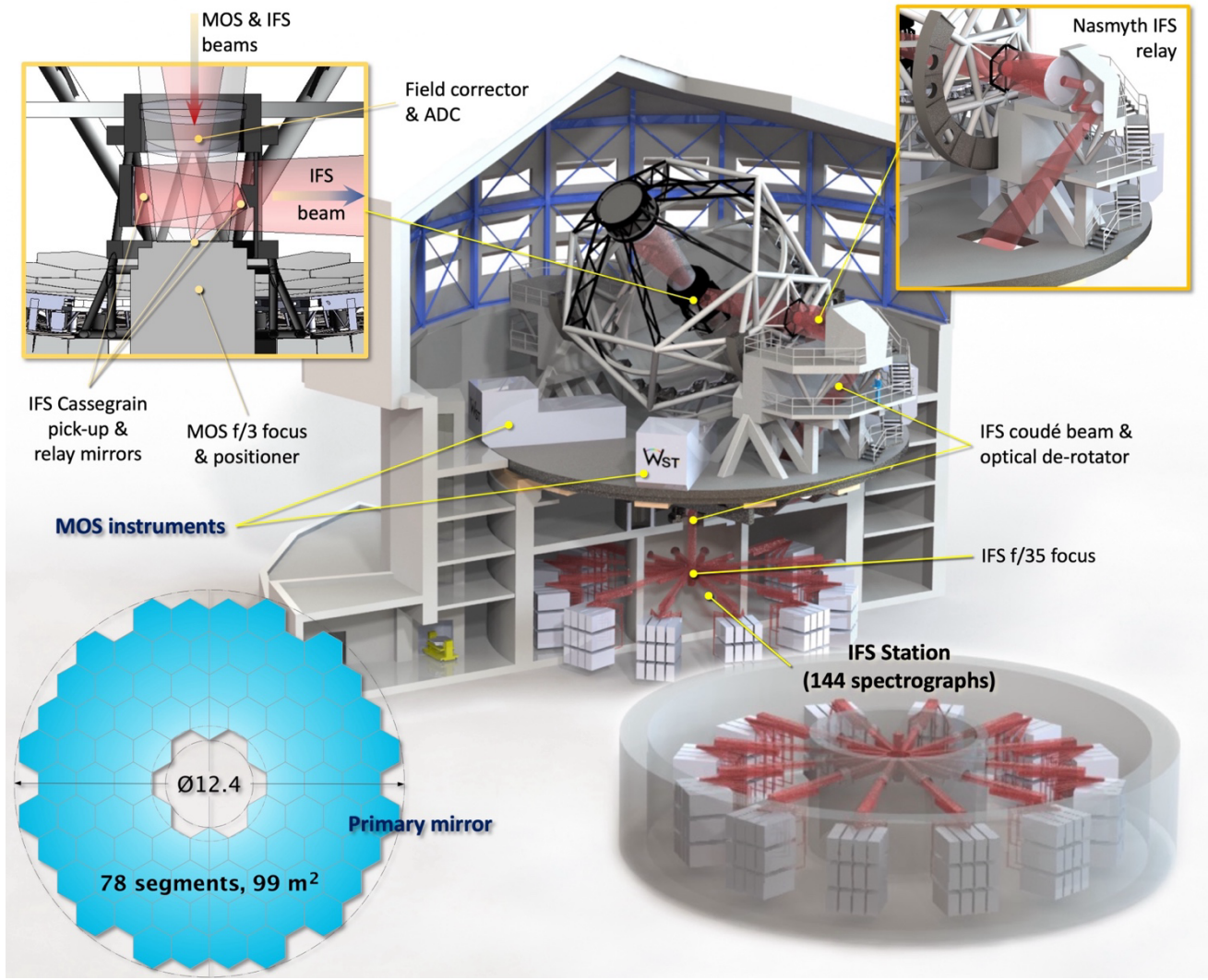

Figure 3 – WST overall mechanical implementation, baseline Cassegrain solution, notional design

The enclosure consists of a multi-level building with room to house the instrumentation at two critical locations: the telescope rotating platform to minimize the length of the MOS fibers and minimize losses, and the lower floor of the enclosure where a large central room is dedicated to the IFS station which requires a controlled environment along with sufficient handling access to operate and maintain the 144 spectrographs. Below the telescope's rotating platform, the pier is hollowed out to host the coudé beam sent from a relay on the Nasmyth platform. Along the vertical path down to the IFS instrument, the 3 x 3 arc mintues$^2$ IFS field is propagated through the de-rotator installed on a dedicated floor. In view of the high telescope pier, particular attention will need to be paid to seismic isolation.

The altitude structure accommodates a foldable bridge crane (Figure 4), for segment replacement. This crane could also hold a local coherencer (LOCO), à la ELT[9]. Its purpose is to ensure that a newly integrated segment is positioned within half a fringe with respect to the neighbouring ones, thereby speeding up on-sky phasing runs[10] and minimizing technical time overheads. We anticipate that freshly re-coated segments would be integrated at a rate of one every 1 to 2 weeks. Assuming compatible interfaces with ELT infrastructure, WST segment recoating would imply a ~10% additional load on the ELT infrastructure.

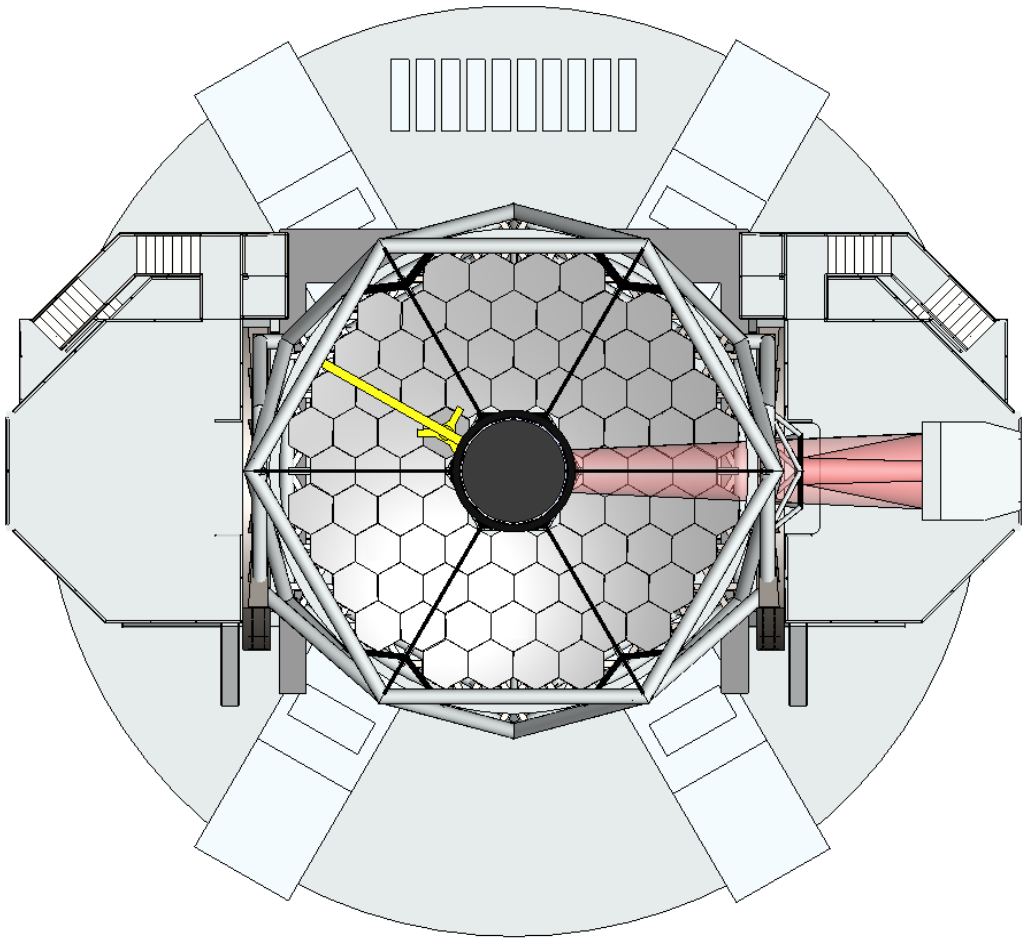

Figure 4 - WST top view, showing the location of the MOS spectrographs. The bridge crane is deployed (yellow strip).

The dome (Figure 3) is designed to minimize the footprint of the building which has a total height of 37 m and a diameter of 38 m – roughly 30% larger than the enclosure of an ESO Very Large Telescope, linear dimensions. The dome is equipped with louvers around its circumference to optimize airflow inside the structure. The main dome shutter splits open in two halves to enable a 13 m wide aperture for the telescope from zenith down to 70 degrees zenithal distance.

## CONCLUSION

WST first baseline design complies with requirements, except for the size of the tip-tilt secondary mirror and vignetting at the IFS focus. With the former, the ongoing iteration restores compliance, thereby lowering the underlying technological risks to manageable levels. This iteration also eliminates the vignetting issue in the IFS relay, improves as-designed optical quality – allowing more margins for error budget, and reduces the difficulty of producing the aspheric lenses of the refracting corrector. The design update still needs to be completed to reduce optical dimensions in the coudé path and thus allow for efficient multi-layer coatings, but we do not anticipate any new non-compliance nor any significant change of the overall layout. Since the IFS and MOS beams share a common path to the Cassegrain-located MOS focal surface, a robust wavefront control scheme driven from the MOS focus and supplemented by internal, IFS-specific controls seems possible. The design foresees a VLT-like enclosure, about 30% larger (linear dimensions).

Acknowledgments: RB, PD, CC thanks CNRS INSU CSAA and ANR-AA-MRSE-2023 for their support. Australia is currently a strategic partner in the European Southern Observatory, supported by the Australian government's 2017-18 Budget measure *Maintaining Australia's Optical Astronomy Capability*."